\documentclass[oldversion]{aa}

\usepackage{graphicx}

\usepackage{txfonts}

\usepackage{color}

\usepackage{natbib}

\usepackage{color}

\newcommand{\qm}[1]{``#1''}

\def\sss{\scriptscriptstyle}
\def\U{{\sss \!U}}
\def\L{{\sss \!L}}
\def\K{{\sss \!K}}

\def\nur{\nu_\mathrm{r}}
\def\nuv{\nu_\theta}
\def\nuL{\nu_\L}
\def\nuU{\nu_\U}
\def\nuK{\nu_\K}

\def\nul{\nuL}
\def\nuu{\nuU}

\definecolor{gray}{rgb}{.6,.6,.6}
\definecolor{green}{rgb}{0,.6,0}
\definecolor{red}{rgb}{.9,0,0}

\begin{document}

\title{ {
Super-spinning compact objects and models of high-frequency quasi-periodic oscillations observed in Galactic microquasars
}}

\author
{Andrea Kotrlov\'a, Gabriel T\"or\"ok, Eva \v{S}r\'amkov\'a \& Zden\v{e}k Stuchl\'{\i}k
}

\institute{
Institute of Physics, Faculty of Philosophy and Science, Silesian
  University in Opava, Bezru\v{c}ovo n\'{a}m. 13, CZ-74601 Opava, Czech Republic
 }
  
\date{Received / Accepted}

\authorrunning{A. Kotrlov\'a et al.}
\titlerunning{Superspinars and QPOs}
 
\date{Received / Accepted}

\abstract
{We have previously applied several models of high-frequency quasi-periodic oscillations (HF QPOs) to estimate the spin of {the} central Kerr black hole in {the} three Galactic microquasars, GRS 1915+105, GRO J1655-40, and XTE J1550-564. Here we explore the alternative possibility that the central compact body is a super-spinning object (or a naked singularity) with the external space-time described by Kerr geometry with a dimensionless spin parameter $a\equiv cJ/GM^2>1$. {We calculate the relevant spin intervals for a subset of HF QPO models considered in the previous study}. Our analysis indicates that for all but one of the considered models there exists at least one interval of $a>1$ that is compatible with constraints given by the ranges of {the central compact object} mass independently estimated for the three sources. For most of the models, the inferred values of $a$ are several times higher than the extreme Kerr black hole value $a=1$. 
These values may be too high since the spin of superspinars is often assumed to rapidly decrease due to accretion when $a\gg1$. In this context, we conclude that only the epicyclic and the Keplerian resonance model provides estimates that are compatible with the expectation of just a small deviation from $a=1$.}


\maketitle
 
\section{Introduction}
\label{section:introduction}

In recent years, several authors have attempted to explore {the limits on various parameters} of accreting compact objects through the phenomenon of high-frequency quasi-periodic oscillations (HF QPOs) and the associated models. These oscillations with frequencies matching those of orbital motion have been observed in the X-ray power density spectra of many black hole (BH) and neutron star (NS) low-mass X-ray binaries \citep[e.g.][]{kli:2006,mcc-rem:2006}. In the NS sources, the HF QPOs usually appear in pairs of two distinct peaks and are sometimes called \qm{twin-peak} HF QPOs. The frequency ratio sampled from their available observations often clusters (typically around 3\,:\,2 value). This can be explained by either the weakness of the two QPOs outside the limited frequency range, the incomplete data sampling, or the intrinsic source clustering \citep[][]{abr-etal:2003, bel-etal:2005,bel-etal:2007a,bel-etal:2007b, tor-etal:2008a, tor-etal:2008b, tor-etal:2008c, tor:2009, bar-bou:2008, bou-etal:2010, wan-etal:2014}. In the BH systems, the HF QPOs appear at frequencies that often form rational ratios with a preferred ratio of 3\,:\,2 \citep[][]{abr-klu:2001,mcc-rem:2003,mcc-rem:2006}. 

Among the varied family of the proposed black hole HF QPO models, there are many of those that attribute the observed {phenomenon} to a certain manifestation of orbital motion of matter orbiting in the accretion disc that surrounds the central compact object. In these models, the observed QPO frequencies are in different ways associated with frequencies of orbital and oscillatory motion of the accretion disc matter. In Kerr geometry, which describes the space-time geometry around a rotating black hole, the relevant frequencies are for geodesic motion dependent solely on the mass and spin of the central compact object. For a particular QPO model and source with an independently estimated mass that displays HF QPOs, it is then possible to calculate the spin of the central object as predicted by the concrete model. These predictions have been {frequently} carried out for three Galactic microquasars with 3\,:\,2 HF QPOs {(GRS 1915+105, GRO J1655-40, XTE J1550-564)} and {various} QPO models \citep[e.g.][]{wag-etal:2001,abr-klu:2001,kat:2004a,kat:2004b,tor-etal:2005,tor-etal:2011,wag:2012,mot-etal:2014a,mot-etal:2014b,stu-kol:2014}. 

For most of the previous estimations, standard Kerr geometry with dimensionless spin parameter $a=cJ/GM^2<1$ was assumed to describe the external space-time around the microquasar. Although commonly accepted, the concept of a Kerr black hole with $a<1$ is at present not the only valuable hypothesis. For instance, consideration of a hypothetical tidal charge implied by the theory of multi-dimensional black holes in the Randall--Sundrum braneworld with non-compactified additional space dimension represents an alternative approach. This approach has been considered in the context of HF QPOs in works of \cite{kot-etal:2008} and \cite{stu-kot:2009}, assuming the Reissner--Nordstr\"{o}m and Kerr--Newman solutions of the Einstein field equations \citep{ali-gum:2005,sch-stu:2009,ali-tal:2009}. Studies dealing with astrophysical applications of generalized non-Kerr space-times have recently received greater attention in a similar context \citep[see e.g.][and references therein]{psa-etal:2008,joh-psa:2011,bam:2012,bam:2014}. Last but not least, Kerr space-time description extended above the black hole limit $a=1$ can be of astrophysical importance here as well \citep[see e.g.][and references therein]{tor-stu:2005,gim-hor:2009, bam-fre:2009,stu-sch:2010,stu-sch:2012a,stu-sch:2012b,stu-sch:2013,bam:2011,li-bam:2013a,li-bam:2013b,kol-stu:2013}.

\begin{table}[t]
\centering
\caption{Properties of the three microquasars GRO 1655-40, GRS 1915+105, and XTE 1550-564.  The individual columns display the frequencies of the lower and upper {$3\!:\!2$} QPO peaks \citep{str:2001,rem-etal:2002,rem-etal:2003} and the estimated ranges of mass \citep{gre-etal:2001, grn-etal:2001, oro-etal:2002, mcc-rem:2003}.}
\label{table:1}
\renewcommand{\arraystretch}{1.2}
\begin{tabular}{lcccc}\\ \hline \hline 
Source & $\nu_{\mathrm{L}}$ [Hz] & $\nu_{\mathrm{U}}$ [Hz] & Mass [$M_{\odot}$] & \\ \hline 
GRO 1655-40 & 300$\pm5$ &  450$\pm3$ & $\phantom{10}$6.0--6.6 & \\ \hline
GRS 1915+105 & 113$\pm5$ &  168$\pm3$ & 10.0--18.0 & \\ \hline
XTE 1550-564 & 184$\pm5$ &  276$\pm3$ & $\phantom{1}$8.4--10.8 &  \\ \hline
\end{tabular}
\end{table}

In this work, we pursue the possibility that the central part of the source comprises a super-spinning compact object (or naked singularity) with the external space-time described by Kerr geometry with dimensionless spin parameter $a>1$.  Following the previous study of \cite{tor-etal:2011}, we explore the predictions of intervals of $a$ that are implied for the three microquasars by several  models of 3\,:\,2 QPOs. The observed QPO frequencies, $\nul$ and $\nuu$,  which we consider next along with the predicted values of mass of the three sources are recalled in Table~\ref{table:1}.

\begin{table}
\caption{{Frequency relations corresponding to individual QPO models. The relations are expressed in terms of three fundamental frequencies of perturbed circular geodesic motion. These are the Keplerian frequency, and the radial and vertical epicyclic frequencies, which are denoted by $\nuK$, $\nur$ and $\nuv$, respectively.}}
\label{table:2}
\renewcommand{\arraystretch}{1.2}
\begin{center}
\begin{tabular}{lll}
    \hline
  \hline
     \textbf{Model} & \multicolumn{2}{c}{\textbf{Relations}} \\
    \hline \hline
$\mathbf{RP}$ & $\nu_{\mathrm{L}} =\nu_{\mathrm{K}}-\nu_{\mathrm{r}}$ & $\nu_{\mathrm{U}} =\nu_{\mathrm{K}}$  \\
$\mathbf{TD}$ & $\nu_{\mathrm{L}}=\nu_{\mathrm{K}}$ & $\nu_{\mathrm{U}}=\nu_{\mathrm{K}}+\nu_{\mathrm{r}}$ \\
\hline
$\mathbf{WD}$ & $\nu_{\mathrm{L}}=2\left(\nu_{\mathrm{K}}-\nu_{\mathrm{r}}\right)$ & $\nu_{\mathrm{U}}=2\nu_{\mathrm{K}}-\nu_{\mathrm{r}}$ \\
$\mathbf{Ep}$ & $\nu_{\mathrm{L}}=\nu_{\mathrm{r}}$ & $\nu_{\mathrm{U}}=\nu_{\theta}$ \\
$\mathbf{Kep}$ & $\nu_{\mathrm{L}}=\nu_{\mathrm{r}}$ & $\nu_{\mathrm{U}}=\nu_{\mathrm{K}}$ \\
$\mathbf{RP1}$ &  $\nu_{\mathrm{L}}=\nu_{\mathrm{K}}-\nu_{\mathrm{r}}$ &  $\nu_{\mathrm{U}}= \nu_{\theta}$ \\
$\mathbf{RP2}$ &  $\nu_{\mathrm{L}}=\nu_{\mathrm{K}}-\nu_{\mathrm{r}}$ &  $\nu_{\mathrm{U}}= 2\nu_{\mathrm{K}}-\nu_{\theta}$ \\
\hline \hline
\end{tabular}
\end{center}
\end{table}

\section{QPO models under consideration}
\label{section:models}

We assume in this paper a subset of QPO models from the group of models considered in \cite{tor-etal:2011} where a detailed description of the models is presented. Here we only briefly list these models and provide the reader with their basic characterization. For more details see \cite{tor-etal:2011} and references therein. The considered models and corresponding frequency relations that determine the expected lower and upper QPO frequency are listed in Table~\ref{table:2}. Frequency relations in the Table are expressed in terms of frequencies of perturbed circular geodesic motion.

Table 2 comprises two models that deal with orbital motion of blobs of matter orbiting in the accretion disc. The "relativistic precession" (RP) model proposed by \cite{ste-vie:1998,ste-vie:1999,ste-vie:2002} attributes the HF QPOs to modes of relativistic epicyclic motion of blobs at various radii in the inner parts of the accretion disc. A similar approach is applied in the so-called "tidal disruption" (TD) model introduced by \cite{cad-etal:2008} and \cite{kos-etal:2009}, in which the QPOs are explained as a manifestation of tidal disruption of large accreting inhomogeneities.

\begin{figure}[t!]
\begin{center}
\includegraphics[width=1\hsize]{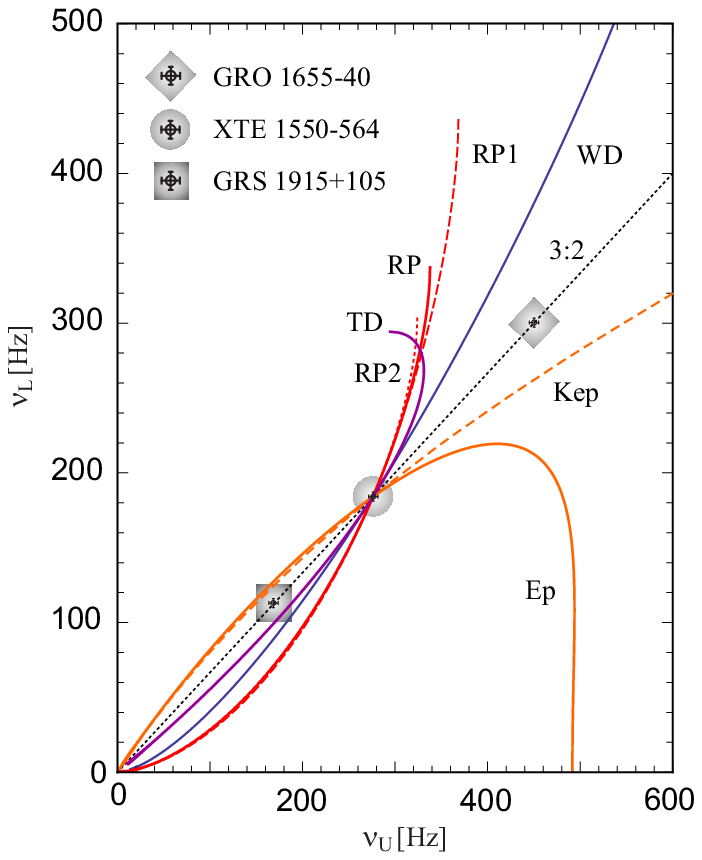}
\end{center}
\caption{{The 3\,:\,2 QPOs of the three microquasars and the particular frequency relations expected for the individual QPO models.  The black dotted line denotes the 3:2 frequency ratio. The {QPO} frequencies and their error bars correspond to those specified in Table~\ref{table:1}. All {expected} frequency relations are plotted for a common referential mass, $M=10M_{\sun}$. The assumed values of spin then scale the curves such that they intersect with the 3:2 relation at frequencies exhibited by XTE~1550-564. Except for the Kep model, these spin values correspond to Kerr black holes, namely to $a=0.468$ for the RP model, to $a = 0.963$ for the Ep model, to $a = 0.339$ for the TD and WD models, to $a=0.662$ for the RP1 model, and to $a=0.372$ for the RP2 model. For the Kep model, the chosen referential value of mass requires the spin $a=1.13$, which exceeds the Kerr black hole limit. }}
\label{figure:1}
\end{figure}

Furthermore, there are several models that are based on the concept of resonance between some modes of accretion disc oscillations. These include the "warped disc" (WD) model introduced by \cite{kat:2004a,kat:2004b}, which, generally speaking, deals with oscillatory modes in a warped accretion disc. Another family of models of this kind is associated with the "resonance" QPO model proposed by \cite{klu-abr:2001}. In this model, the HF QPOs are ascribed to a non-linear resonance between two modes of accretion disc oscillations. From various alternatives of the resonance model we consider here the two most popular. These are represented by the "epicyclic resonance" (Ep) model dealing with radial and vertical epicyclic oscillations, and the "Keplerian resonance" (Kp) model assuming a resonance between the orbital Keplerian motion and the radial epicyclic oscillations. Finally, there are another two resonance models that consider different combinations of non-axisymmetric disc-oscillation modes. We refer to them as the "RP1" model \citep{bur:2005} and the "RP2" model \citep{tor-etal:2010}. 

Frequency relations predicted by the above recalled models are sensitive to parameters of the central compact object. Their behaviour is illustrated in Figure~\ref{figure:1} together with the 3\,:\,2  QPOs displayed by the three microquasars.

\begin{figure*}[t]
\begin{center}
\includegraphics[width=1\hsize]{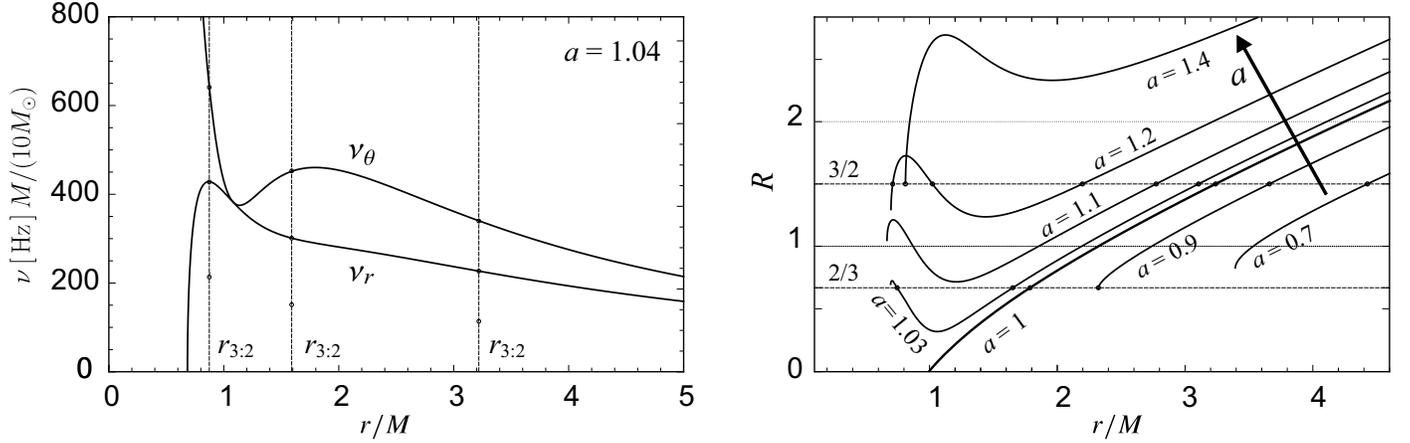}
\end{center}
\caption{Left: Behaviour of the epicyclic frequencies for $a=1.04$. The three possible QPO excitation radii determined within the Ep model are denoted by the dotted vertical lines.  We note that more than one extremum of the epicyclic frequencies related to accretion discs appear only when $a\in(0.952,~1.089)$. Right: Frequency ratio $R_{\mathrm{RP1}}$ determined within the RP1 model for various values of $a$. We note that  the observed 3\,:\,2 ratio is not allowed by the model when $a>1.4$.}
\label{figure:2}
\end{figure*}

\section{Spin intervals implied by the QPO models}
\label{section:intervals}

The frequency relations of the considered QPO models, which are specified in Table~\ref{table:2} and illustrated in Figure~\ref{figure:1}, are given by the orbital Keplerian frequency $\nu_{\mathrm{K}}$ and the radial $\nu_{\mathrm{r}}$ and vertical $\nu_{\theta}$ epicyclic frequencies. In Kerr geometry, these are all functions solely of mass $M$ and spin $a$ of the central compact object. The explicit formulae of these functions have been derived by \cite{ali-gal:1981} for Kerr black holes and also investigated by \cite{oka-etal:1987,now-leh:1999} and others. More recently, \cite{tor-stu:2005} explored the behaviour of the three fundamental frequencies for the case when $a>1$, which is associated with Kerr naked singularities or superspinning compact objects. A detailed discussion of the behaviour of functions, which determine the QPO excitation radii for different combinations of disc oscillation modes, was presented in \cite{tor-etal:2005, tor-stu:2005,stu-sch:2012b} and \cite{stu-etal:2013}. Armed with the previously achieved results, we can calculate values of the compact object spin as predicted by the considered QPO models along with the observed QPO frequencies assuming that $a\in(0,\,\infty)$.\footnote{It should be noted that the HF QPOs are well correlated with the low-frequency QPO features \citep[e.g.][]{psa-etal:1999,kli:2006,bel-ste:2014}. Some of the models considered here indeed assume such correlations, e.g., the Lense-Thirring precession within the framework of the RP model \cite[][]{ste-vie:1999, ste-etal:1999}, or the so-called 13th wave within the framework of the Ep model \citep{abr-etal:2004}. It should be thus possible to extend our consideration to low-frequency QPOs \citep[see also][]{mot-etal:2014a,mot-etal:2014b}. Nevertheless,  we focus here in a coherent way on a larger set of models restricting our attention entirely to HF QPOs.}  

\subsection{Excitation radii and spin functions}
\label{section:spin-functions}

As discussed in detail by \cite{tor-stu:2005}, the behaviour of the epicyclic frequencies is most complicated for $a\gtrsim 1$, exhibiting large differences from the cases of $0\leq a\leq1$ and $a\gg1$. This has a direct impact on the location of the QPO excitation radii $r_{3:2}$ identified within a given model. In the left panel of Figure~\ref{figure:2}, we plot the radial and vertical epicyclic frequencies, $\nur$ and $\nuv$, for {$a=1.04$}. We can see that for the Ep model there are three possible QPO excitation radii. This creates a more complex situation than in the previously examined black hole case, in which only one such radius appeared. In general, the relevant spin function determined for the 3\,:\,2 frequency ratio, which occurs at the resonant radius $x_{3:2}=r_{3:2}/M$, has two separate branches of solution given as follows
$$
a_{\mathrm{Ep}}(x_{3:2})=\frac{\sqrt{x_{3:2}}}{39}\left[44 \mp \sqrt{5\left[39(x_{3:2}) - 34\right]}\right],
$$
where the \qm{$-$} sign applies to black holes and naked singularities with $a \sim 1$, while the \qm{$+$} sign applies to naked singularities with $a>1$.

Assuming the (geodesic) Ep model, the upper QPO frequency must always be assigned to the vertical axisymmetric disc oscillation mode, $\nuu=\nuv$. This is the case since $\nuv\geq\nur$ for any $a$. However, the assigment of $\nuu$ and $\nul$ can be ambiguous within some models. We illustrate this issue in the right panel of Figure~\ref{figure:2}, which shows the frequency ratio $R_{\mathrm{RP1}}\equiv \nuv/(\nuK-\nur)$ predicted by the RP1 model for various radii and spins. We can see that the model allows for the $3/2$, as well as the $2/3$, frequency ratio when $a\in(0.9,\,1.1)$. We note that such behaviour arises due to the existence of the maximum of the vertical epicyclic frequency. We also note that the RP1 model does not reproduce the 3\,:\,2 ratio for any frequency assigment when $a>1.4$ (see the trend  of curves in the right panel of Figure~\ref{figure:2}). 

\begin{figure*}[t]
\begin{center}
\includegraphics[width=0.95\hsize]{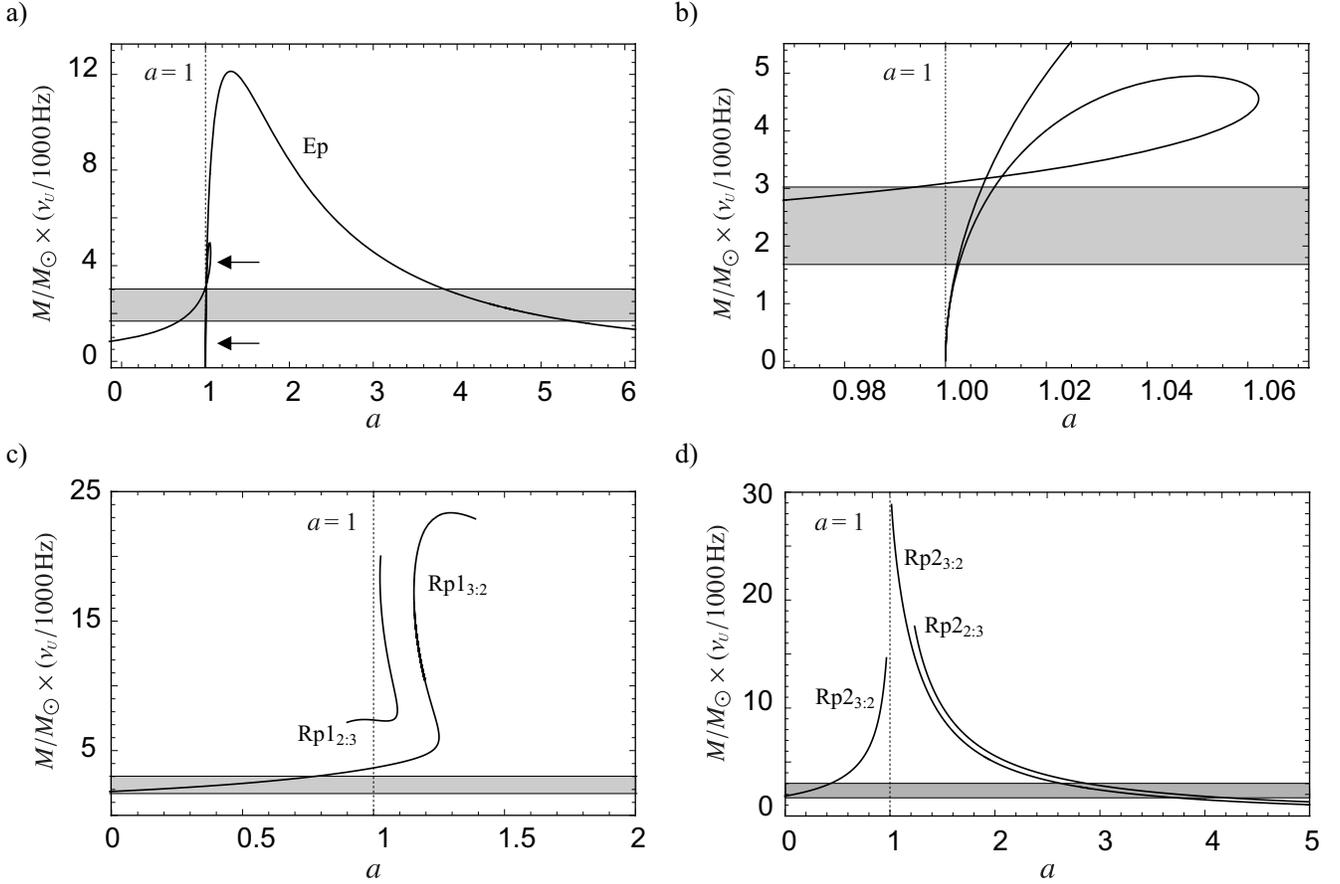}
\end{center}
\caption{Behaviour of the ambiguous $M\times\nuU(a)$ relations discussed in Section~\ref{section:spin-functions}. The shaded rectangular area denotes the observational bounds on the quantity $M\times\nuU(a)$ determined for GRS~1915+105 (Table~\ref{table:1}). This area also roughly indicates the upper bounds for the whole group of the three microquasars GRS~1915$+$105, XTE~J1550$-$564 and GRO~J1655$-$40. a) The $M\times\nuU(a)$ curves drawn for the Ep model. The two arrows indicate a region close to $a=1$ where more than one resonant radii appear. b) An enlarged view of the curves drawn for the Ep model in the region close to $a=1$. c) The $M\times\nuU(a)$ curves drawn for the RP1 model. Note the two branches of the curves and their full absence above $a=1.4$. d) The $M\times\nuU(a)$ curves drawn for the RP2 model. Note the discontinuities in the displayed curves.}
\label{figure:3}
\end{figure*}

The illustrated existence of more than one excitation radii for a given spin, the ambiguities in the frequency assigment, and the intrinsic restrictions to the predicted frequency ratio must be fully taken into account when evaluating appropriate spin functions for a given QPO model. We have investigated all models listed in Table~\ref{table:2} across the whole range of $a\geq0$. We find that the Ep, RP1, and RP2 models imply rather complicated $\nuU(a)$ relations. These relations are displayed in their mass-independent form in Figure~\ref{figure:3}. On the other hand, the RP model together with the WD and Kep models imply simple, although non-monotonic, continuous functions $\nuU(a)$. These functions are included in the mass-independent form in Figure~\ref{figure:4}. 

\subsection{Inferred values of spin}

The observational bounds on the quantity $M\times\nuU(a)$ determined for GRS~1915+105 roughly represent the upper bounds common for all the three microquasars (see Table~\ref{table:1}). We denote them by the shaded area in Figures~\ref{figure:3} and \ref{figure:4}. The mass-independent form of the $M\times\nuU(a)$ relations displayed in these figures is implied by the relativistic $1/M$ scaling of the orbital frequencies.  This allows for a brief, straightforward comparison between the expected and the observed quantities $M$, $a$ and $\nuu$. Detailed information on spin intervals predicted by the individual models for each microquasar is then given in Table~\ref{table:3}.

\begin{table*}
\caption{Spin intervals implied by the individual QPO models for the three microquasars GRS~1915$+$105, XTE~J1550$-$564, and GRO~J1655$-$40.}
\label{table:3}
\renewcommand{\arraystretch}{1.4}
\begin{center}
\begin{tabular}{l|lc|lc|lc}
    \hline
  \multicolumn{7}{c}{$a$}\\\hline
     \textbf{Model} & \multicolumn{2}{c}{\textbf{GRS~1915$+$105}} & \multicolumn{2}{|c|}{\textbf{XTE~J1550$-$564}} & \multicolumn{2}{c}{\textbf{GRO~J1655$-$40}}\\
    \hline \hline
$\mathbf{RP}$  & $<0.55$ & $4.17 - 5.99$ & $0.29 - 0.54$ & $4.21 - 4.92$ & $0.45 - 0.53$ & $4.22 - 4.48$\\
\hline
$\mathbf{TD}$  & $<0.44$ & $5.17 - 7.34$ & $0.12 - 0.43$ & $5.22 - 6.06$ & $0.31 - 0.42$ & $5.23 - 5.54$\\
\hline
$\mathbf{WD}$  & $<0.44$ & $5.17 - 7.34$ & $0.12 - 0.43$ & $5.22 - 6.06$ & $0.31 - 0.42$ & $5.23 - 5.54$ \\
\hline
$\mathbf{Ep}$  & $0.68 - 0.99$ &         & $0.89 - 0.99$ &               & $0.96 - 0.99$ &  \\
               & $1.00229 - 1.00729$ &   & $1.00431 - 1.00708$ &         & $1.00582 - 1.00703$ & \\
               & $1.00264 - 1.00972$ & $3.84 - 5.38$ & $1.00530 - 1.00939$ & $3.88 - 4.48$ & $1.00746 - 1.00931$ & $3.88 - 4.11$\\
\hline
$\mathbf{Kep}$ & $0.79 - 1.17$ & $3.81 - 5.41$ & $1.03 - 1.17$ & $3.84 - 4.47$ & $1.12 - 1.16$ & $3.85 - 4.08$\\

\hline
$\mathbf{RP1}$ & $<0.78$ & $-$ & $0.41 - 0.76$ & $-$ & $0.63 - 0.76$ & $-$ \\
\hline
$\mathbf{RP2}$ & $<0.44$ & $2.63 - 4.21$ & $0.23 - 0.43$ & $2.65 - 3.40$ & $0.35 - 0.43$ & $2.65 - 3.08$\\
\hline \hline
\end{tabular}
\end{center}
\end{table*}

\section{Discussion and conclusions}
\label{section:conclusions}

Numerous attempts to estimate the spin $a$ of black hole candidates have been undertaken in the past decade using the iron-line profile, the X-ray continuum, or the QPO frequency fitting-methods. So far the most promising of these studies seem to be the applications of the X-ray continuum fitting method \citep[e.g.][]{mcc-etal:2011}. In some cases there is a good agreement between the particular applications of the two former methods mentioned above \cite[][]{ste-etal:2011}. However, there remain discrepancies between the results of some of the studies and no agreement between all the three approaches has been yet achieved \citep[see also discussion in][]{tor-etal:2011}. The attempts that consider Kerr black hole space-times are by definition limited to $a<1$. In principle, one can consider the extension of the two X-ray spectral fitting methods above the $a=1$ limit \citep[e.g.][]{tak-har:2010,sch-stu:2013}. Rather natural is then also analogic extension of the X-ray timing approach which incorporates the same super-spinning compact objects with the exterior described by the space-time of a Kerr naked singularity and the interior given by solutions of the string theory.  

\begin{figure}
\begin{center}
\includegraphics[width=.96\hsize]{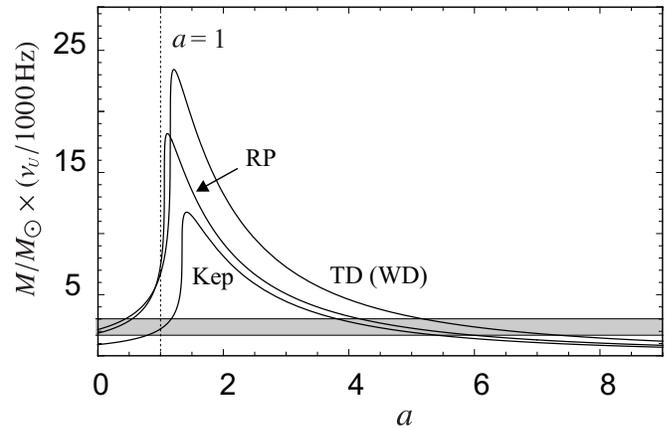}
\end{center}
\caption{Behaviour of the $M\times\nuU(a)$ functions in the case of the RP, Kep, TD, and WD models.  The shaded rectangular area denotes the observational bounds on the quantity $M\times\nuU(a)$ determined for GRS~1915+105 (Table~\ref{table:1}). This area also roughly indicates the upper bounds for the whole group of the three microquasars GRS~1915$+$105, XTE~J1550$-$564, and GRO~J1655$-$40. Note that the curve corresponding to the TD model joins those that correspond to the WD model. The coincidence follows from the model frequency relations and the requirement of the observed 3\,:\,2 frequency ratio.}
\label{figure:4}
\end{figure}

From our results, illustrated in Figures ~\ref{figure:3} and \ref{figure:4}, one can conclude that, except for the RP1 model, all the considered models enter at least one interval of $a>1$ that is compatible with the observational constraints on mass and the QPO frequency. In more detail, all these intervals are located below the value of $a=8$. The expectation of $a>1$ therefore provides an alternative for almost all of the QPO models considered here. We question, however, whether or not most of the values of $a>1$ given in Table~\ref{table:3} are too high. It has been demonstrated that angular momentum of superspinars should be continuously lowered \citep[e.g.][]{cal-nob:1979,stu:1980,stu:1981,stu-etal:2011}. Their spin tends to rapidly decrease due to accretion for $a\gg1$, while it is less affected for $a\gtrsim1$. In this context, we emphasize the finding that only the epicyclic and similar Keplerian resonance models comply with the expectation of a small deviation of $a$ from the value $a=1$. 

Finally, we note that for the epicyclic resonance model there is the possibility of recognizing a direct observational signature of presence of a super-spinning compact object. It is clear from panels a) of Figure~\ref{figure:2} and b) of Figure~\ref{figure:3} that more pairs of different 3:2 commensurable frequencies can be expected within a single source. Further treatment of this issue is rather difficult considering the present lack and low-resolution of the BH HF QPO data. It should, however, be resolvable using the large amount of high quality data available through the next generation of X-ray observatories, such as the proposed Large Observatory for X-ray Timing \citep[LOFT;][]{fer-etal:2012}.

\section*{Acknowledgments}
GT and ES would like to acknowledge the Czech grant GA\v{C}R 209/12/P740. ZS acknowledges the Albert Einstein Centre for Gravitation and Astrophysics supported by the Czech Science Foundation grant No. {14-37086G}. Furthermore, we acknowledge the grant 02983/2013/RRC conducted in the framework of the funding programme \qm{Research and Development Support in Moravian-Silesian Region}.

\bigskip


\end{document}